\documentstyle[epsf]{elsart}
\newcommand\di{_{di}}
\newcommand\dmmi{_{d-2,i}}
\newcommand\dmi{_{d-1,i}}
\newcommand\dpi{_{d+1,i}}
\newcommand{\slu}[1]{\sigma_{\scriptscriptstyle #1}}
\newcommand{\EL}[1]{E_{\scriptscriptstyle #1}}
\newcommand{\GL}[1]{G_{\scriptscriptstyle #1}}
\newcommand{\gl}[1]{g_{\scriptscriptstyle #1}}
\newcommand{\ai}{a_{\scriptscriptstyle s}}
\newcommand{\flu}[1]{f_{\scriptscriptstyle #1}}
\newcommand{\Glu}{\Gamma_{\scriptscriptstyle LU}}
\newcommand{\Clu}{C_{\scriptscriptstyle LU}}
\newcommand{\Elu}{E_{\scriptscriptstyle LU}}
\newcommand{\nulu}[1]{\nu_{\scriptscriptstyle #1}}

\newcommand{\Teff}{\hbox{$T_{\rm eff}$}}

\def\logg{$\log g$\thinspace}

\begin{document}
\begin{frontmatter}
\title{The Classical Stellar Atmosphere Problem}
\author{Klaus Werner}
\author{and Stefan Dreizler}
\address{Institut f\"ur Astronomie und Astrophysik, Universit\"at T\"ubingen,
\\ Waldh\"auser Str. 64, D-72076 T\"ubingen, Germany}
\begin{abstract}
We introduce the classical stellar atmosphere problem and describe in detail
its numerical solution. The problem consists of the solution of the radiation
transfer equations under the constraints of hydrostatic, radiative and
statistical equilibrium (non-LTE). We outline the basic idea of the Accelerated
Lambda Iteration (ALI) technique and statistical methods which finally allow the
construction of non-LTE model atmospheres considering the influence of
millions of metal absorption lines. Some applications of the new models are
presented.
\end{abstract}
\end{frontmatter}

\section{Introduction}

The quantitative analysis of stellar spectra is one of the most important tools
of modern astrophysics. Basically all our knowledge about structure and
evolution of stars, and hence about galactic evolution in general, rests on the
interpretation of their electromagnetic spectrum. The formation of the observed
spectrum is usually confined to a very thin layer on top of the stellar core,
the atmosphere. Spectral analysis is performed by modeling the temperature and
pressure stratification of the atmosphere and computing synthetic spectra which
are then compared to observation. Fitting synthetic spectra from a grid of
models yields the basic photospheric parameters, effective temperature,
surface gravity, and chemical composition. Comparison with theoretical
evolutionary calculations allows the derivation of stellar parameters like
mass, radius and total luminosity. 

The so-called classical stellar atmosphere problem considers the transfer of
electromagnetic radiation, released by interior energy sources, through the
outermost layers of a star into free space by making three specific physical
assumptions. At first it is assumed that the atmosphere is in hydrostatic
equilibrium, thus, the matter which interacts with photons is at rest. Second,
the transfer of energy through the atmosphere is entirely due to photons, i.e.\
heat conduction and large scale convection are regarded as negligible
(so-called radiative equilibrium). The effectiveness of photon transfer depends
on the total opacity and emissivity of the matter which are strongly state and
frequency dependent quantities. They  depend in detail on the occupation
density of atomic levels which in turn are determined by the local temperature
and electron density as well as by the radiation field, whose nature is
non-local in character. The occupation of any atomic level is balanced by
radiative and collisional population and de-population processes (statistical
equilibrium; our third assumption), i.e.\ the interaction of atoms with other
particles and photons.  Mathematically, the whole problem consists of the
solution of the radiation transfer equations simultaneously with the equations
for hydrostatic and radiative equilibrium, together with the statistical
equilibrium, or, rate equations.

A stellar atmosphere is radiating into the circumstellar space and thus
evidently is an open thermodynamic system, hence it cannot be in thermodynamic
equilibrium (TE) and thus we cannot simply assign a temperature. The ``Local
Thermodynamic Equilibrium'' (LTE) is  a working hypothesis which assumes TE not
for the atmosphere as a whole but for small volume elements. As a consequence,
the atomic population numbers are depending only on the local (electron)
temperature and electron density via the Saha-Boltzmann equations. Computing
models by replacing the Saha-Boltzmann equations by the rate equations are
called non-LTE (or NLTE) models. This designation is unfortunate because still,
the velocity distribution of particles is assumed to be Maxwellian, i.e.\ we
can still define a local temperature. NLTE calculations are tremendously more
costly than LTE calculations, however, it is hard to predict if NLTE effects
are important in a specific problem. Generally, NLTE effects are large at high
temperatures and low densities, which implies intense radiation fields hence
frequent radiative processes and less frequent particle collisions which tend
to enforce LTE conditions.

Relaxing the LTE assumption leads to the classical model atmosphere problem,
i.e.\ solution of the radiation transfer equations assuming hydrostatic,
radiative and statistical equilibrium. Such models are applicable to the vast
majority of stars. The numerical problem going from LTE to realistic NLTE
models has only recently been solved and is the topic of this paper. We now
have the tools in hand to consider non-classical models, which consider the
radiation transfer in more general environments, for example in expanding
stellar atmospheres. This is the topic of another paper in this volume
\cite{HB98}.

Stellar atmosphere modeling has made significant progress within the recent
years. This is based on the development of new numerical techniques for model
construction as well as on the fact that reliable atomic data have become
available for many species. Of course these achievements go along with a strong
increase of computing power. Model atmospheres assuming LTE have been highly
refined by the inclusion of many more atomic and molecular opacity sources,
however, elaborated numerical techniques for LTE model computation are
available for many years. The progress is most remarkable in the field of NLTE
model atmospheres. The replacement of the Saha-Boltzmann equations (LTE) by the
atomic rate equations (NLTE) requires a different numerical solution technique,
otherwise metal opacities cannot be accounted for at all. Such techniques were
developed with big success during the last decade, triggered by important
papers by Cannon \cite{Can73} and Scharmer \cite{Schar81}. The Accelerated Lambda
Iteration (ALI) is the basis of this development. Combined with statistical
methods we are finally able to compute so-called metal line blanketed NLTE
models (considering many millions of spectral lines) with a very high level of
sophistication.

In this paper we discuss the basic ideas behind the new numerical methods for
NLTE modeling. At first we state the classical model atmosphere problem and
describe the ALI solution technique. We then focus on the NLTE metal line
blanketing problem and its solution by the introduction of the superlevel
concept and statistical methods to treat the opacities (Opacity Sampling and
Opacity Distribution Functions). Finally we demonstrate successful applications
of the new models by presenting a few exemplary case studies.

\section{Statement of the problem and overview of the solution method\label{eins2}}

In the following text we outline the general stellar atmosphere problem, but
will discuss various details of numerical implementation as applied to our
computer program {\tt PRO2}. We assume plane parallel geometry, which is well
justified for most stars because the atmospheres are thin compared to the
stellar radius. The only parameters which characterize uniquely such an
atmosphere are the effective temperature (\Teff), which is a measure for the
amount of energy transported through the atmosphere per unit area and time (see
Eq.\,\ref{nominal}), the surface gravity ($g$), and the chemical composition.
Generalization to spherical symmetry to account for extended (static)
atmospheres mainly affects the radiation transfer equation and is
straightforward \cite[p.\,250]{Mi78}. To construct model atmospheres we have
developed our program which solves simultaneously a set of equations that is
highly coupled and non-linear. Because of the coupling, no equation is
determining uniquely a single quantity -- all equations determine a number of
state parameters. However, each of them is usually thought of as determining a
particular quantity. These equations are:

\begin{itemize}
\item The radiation transfer equations which are solved for the (angular) mean
intensities $J_i, i=1,\ldots,NF$, on a pre-chosen frequency grid comprising
$NF$ points. The formal solution is given by $J=\Lambda S$, where $S$ is the
source function as defined later (Eq.\,\ref{source}). Although $\Lambda$ is
written as an operator, one may think of $\Lambda$ as a {\em process} of
obtaining the mean intensity from the source function.
\item The hydrostatic equilibrium equation which determines the total particle
density $N$.
\item The radiative equilibrium equation from which the temperature $T$ follows.
\item The particle conservation equation, determining the electron density 
$n_e$.
\item The statistical equilibrium equations which are solved for the
population densities $n_i, i=1,\ldots,NL$  of the atomic levels allowed to
depart from LTE (NLTE levels).
\item The definition equation for a fictitious massive particle density $n_H$
which is introduced for a convenient representation of the solution procedure.
\end{itemize}

This set of equations has to be solved at each point $d$ of a grid
comprising $ND$ depth points. Thus we are looking for solution vectors
\begin{equation}\label{psi1}
\mbox{\boldmath$\psi$}_d' = (n_1,\ldots,n_{NL}, n_e, T, n_H, N,
J_1,\ldots,J_{NF}) , \quad d=1, \ldots ,ND .
\end{equation}
The Complete Linearization (CL) method \cite{AM69} solves this set by
linearizing the equations with respect to all variables. The basic advantage of
the ALI (or ``operator splitting'') method is that it allows to eliminate at
the outset the explicit occurrence of the mean intensities $J_i$ from the
solution scheme by expressing these variables by the current, yet to be
determined, occupation densities and temperature. This is accomplished by an
iteration procedure which may be written as (suppressing indices indicating
depth and frequency dependency of variables):
\begin{equation}\label{ali}
J^n=\Lambda^{\star}S^n+(\Lambda-\Lambda^{\star})S^{n-1} .
\end{equation}
This means that the actual mean intensity at any iteration step $n$ is computed
by applying an Approximate Lambda Operator (ALO) $\Lambda^{\star}$ on the
actual (thermal) source function $S^n$ plus a correction term that is computed
from quantities known from the previous iteration step. This correction term
includes the exact lambda operator $\Lambda$ which guarantees the exact
solution of the radiation transfer problem in the limit of convergence:
$J=\Lambda S$. The use of $\Lambda$ in Eq.\,\ref{ali} only indicates that a
formal solution of the transfer equation is performed but in fact the operator
is usually not constructed explicitly. Instead a Feautrier solution scheme
\cite[p.\,156]{Mi78} or any other standard method can be employed to solve the
transfer equation that is set up as a differential equation.

The resulting set of equations for the reduced solution vectors
\begin{equation}\label{psi2}
\mbox{\boldmath$\psi$}_d = (n_1,\ldots,n_{NL}, n_e, T, n_H, N) , \quad
d=1,\ldots,ND
\end{equation}
is of course still non-linear. The solution is obtained by linearization and
iteration which is performed either with a usual Newton-Raphson iteration or by
other, much faster methods like the quasi-Newton or Kantorovich variants
\cite{DW91,HL92}. The first model atmosphere calculations with the ALI
method were performed by Werner \cite{W86}.

Another advantage of the ALI method is that the explicit depth coupling of the
solution vectors Eq.\,\ref{psi1} through the transfer equation can be avoided
if one restricts to diagonal (i.e.\ local) approximate $\Lambda$-operators.
Then the solution vectors Eq.\,\ref{psi2} are independent from each other and
the solution procedure within one iteration step of Eq.\,\ref{ali} is much more
straightforward. Depth coupling is provided by the correction term that
involves the exact solution of the transfer equation. The hydrostatic equation
which also gives an explicit depth coupling, may be taken out of the set of
equations and can -- as experience shows -- be solved in between two iteration
steps of Eq.\,\ref{ali}. Then full advantage of a local ALO can be taken.

The linearized system may be written as
\begin{equation}
\mbox{\boldmath$\psi$}_d
=\mbox{\boldmath$\psi$}_d^0+\delta\mbox{\boldmath$\psi$}_d
\end{equation}
where $\mbox{\boldmath$\psi$}_d^0$ is the current estimate for the
solution vector at depth $d$ and $\delta\mbox{\boldmath$\psi$}_d$ is the
correction vector to be computed. Using a tri-diagonal $\Lambda^{\star}$
operator the resulting system for $\delta\mbox{\boldmath$\psi$}_d$ is -- like
in the classical CL scheme -- of block tri-diagonal form coupling each depth
point $d$ to its nearest neighbors $d\pm1$:
\begin{equation}\label{tri}
\mbox{\boldmath$\gamma$}_d\delta\mbox{\boldmath$\psi$}_{d-1}+
\mbox{\boldmath$\beta$}_d\delta\mbox{\boldmath$\psi$}_d+
\mbox{\boldmath$\alpha$}_d\delta\mbox{\boldmath$\psi$}_{d+1}={\bf c}_d .
\end{equation}
The quantities {$\mbox{\boldmath$\alpha, \beta, \gamma$}$} are ($NN\times NN$)
matrices where $NN$ is the total number of physical variables, i.e., $NN=NL+4$,
and ${\bf c}_d$ is the residual error in the equations. The solution is
obtained by the Feautrier scheme. With starting values
${\bf D}_1=\mbox{\boldmath$\beta$}_1^{-1}(\mbox{-\boldmath$\alpha$}_1$) and ${\bf
v}_1=\mbox{\boldmath$\beta$}_1^{-1}{\bf c}_1$ we sweep from the outer boundary
of the atmosphere inside and calculate at each depth:
\begin{eqnarray}\label{feau}
{\bf D}_d&=&(\mbox{\boldmath$\beta$}_d+
\mbox{\boldmath$\gamma$}_d{\bf D}_{d-1})^{-1}
(-\mbox{\boldmath$\alpha$}_d)\nonumber\\
{\bf v}_d&=&
(\mbox{\boldmath$\beta$}_d+\mbox{\boldmath$\gamma$}_d{\bf D}_{d-1})^{-1}
({\bf c}_d-\mbox{\boldmath$\gamma$}_d{\bf v}_{d-1}) .
\end{eqnarray}
At the inner boundary we have $\mbox{\boldmath$D$}_{ND}=\mbox{\bf 0}$ and
sweeping back outside we calculate the correction vectors, first
$\delta\mbox{\boldmath$\psi$}_{ND}={\bf v}_{ND}$ and then successively
$\delta\mbox{\boldmath$\psi$}_d={\bf
D}_d\delta\mbox{\boldmath$\psi$}_{d+1}+{\bf v}_d$. As already mentioned, the
system Eq.\,\ref{tri} breaks into $ND$ independent equations
$\delta\mbox{\boldmath$\psi$}_d=\mbox{\boldmath$\beta$}_d^{-1}{\bf c}_d$
($d=1,\ldots,ND$) when a local $\Lambda^{\star}$ operator is used. The
additional numerical effort to set up the subdiagonal matrices and matrix
multiplications in the tri-diagonal case is outweighed by the faster global
convergence of the ALI cycle, accomplished by the explicit depth coupling in the
linearization procedure \cite{W89}.

The principal advantage of the ALI over the CL method becomes clear at this
point. Each matrix inversion in Eq.\,\ref{feau} requires $(NL+4)^3$ operations
whereas in the CL method $(NL+NF+4)^3$ operations are needed. Since the number
of frequency points $NF$ is much larger than the number of levels $NL$, the
matrix inversion in the CL approach is dominated by $NF$.

Recent developments concern the problem that the total number of atomic levels
tractable in NLTE with the ALI method described so far is restricted to the
order of 250, from our experience with {\tt PRO2}. This limit is a consequence
of the non-linearity of the equations, and in order to overcome it,  measures
must be taken in order to achieve a linear system whose numerical solution is
much more stable. Such a pre-conditioning procedure has been first applied in
the ALI context by Werner \& Husfeld \cite{WH85}. More advanced work achieves
linearity by replacing the $\Lambda$ operator with the $\Psi$ operator (and by
judiciously considering some populations as ``old'' and some as ``new''
ones within an ALI step) which is formally defined by writing
\begin{equation}
J_\nu=\Psi_\nu\eta_\nu\ , \qquad {\rm i.e.} \qquad
\Psi_\nu\equiv\Lambda_\nu/\chi_\nu\ ,
\end{equation}
where the total opacity $\chi_\nu$ (as defined in Sect.\,\ref{opa}) is
calculated from the previous ALI cycle. The advantage is that the emissivity
$\eta_\nu$ (Sect.\,\ref{opa}) is linear in the populations, whereas the source
function $S_\nu$ is not. Hence the new operator $\Psi$ gives the solution of the
transfer problem by acting on a linear function. This idea is based on Rybicki
\& Hummer \cite{RH91} who applied it to the line formation problem, i.e.
restricting the set of equations to the transfer and rate equations and
regarding the atmospheric structure as fixed. Hauschildt \etal
\cite{HB98,Hau93} generalized it to solve the full model atmosphere problem. In
addition, splitting the set of statistical equations and solving it separately
for each chemical element means that now many hundreds of levels per species
are tractable in NLTE. A very robust method and fast variant of the ALI method,
the ALI/CL hybrid scheme, allows for the linearization of the radiation field
for selected frequencies \cite{HL95}, but it is not implemented in {\tt PRO2}.

\section{Basic equations}

\subsection{Radiation transfer}

Any numerical method requires a formal solution (i.e.\ atmospheric structure
already given) of the radiation transfer problem. The radiation transfer at any
particular depth point can be described by the following equation, formally
written for positive and negative $\mu$ (which is the cosine of the angle
between direction of propagation and outward directed normal to the surface)
separately, i.e.\ for inward and for outward directional intensities $I$ with
frequency $\nu$:
\begin{equation}\label{te}
\pm\mu \frac{\partial I_{\nu}(\pm\mu)}{\partial\tau_{\nu}}=
S_{\nu}-I_{\nu}(\pm\mu), \quad\mu\in[0,1].
\end{equation}
$\tau_\nu$ is the optical depth (which can be defined via the column mass $m$
that is used in the other structural equations and later introduced in
Sect.\,\ref{defm} by $d\tau_\nu=dm \chi_\nu/\rho$, with the mass density $\rho$)
and $S_{\nu}$ is the local source function. Introducing the Feautrier variable
\begin{equation}\label{udef}
u_{\nu\mu}\equiv \left(I_{\nu}(\mu)+I_{\nu}(-\mu)\right)/2
\end{equation}
we obtain the second-order form \cite[p.\,151]{Mi78}:
\begin{equation}
\mu^2 \frac{\partial^2u_{\nu\mu}}{\partial\tau_{\nu}^2}=u_{\nu\mu}-S_{\nu}, \quad\mu\in[0,1].
\end{equation}
We may separate the Thomson emissivity term (scattering from free electrons,
assumed coherent, with cross-section $\sigma_e$) from the source function so
that
\begin{equation}\label{source}
S_{\nu}=S_{\nu}'+n_e\sigma_eJ_{\nu}/\chi_{\nu},
\end{equation}
where $S_{\nu}'$ is the ratio of thermal emissivity to total opacity as described in
detail below (Sect.\,\ref{opa}): $S_{\nu}'=\eta_{\nu}/\chi_{\nu}$. Since the mean
intensity is the angular integral over the Feautrier intensity the transfer
equation becomes
\begin{equation}\label{unm}
\mu^2 \frac{\partial^2u_{\nu\mu}}{\partial\tau_{\nu}^2}
=u_{\nu\mu}-S_{\nu}'-\frac{n_e\sigma_e}{\chi_{\nu}}\int_{0}^{1}u_{\nu\mu}\,d\mu .
\end{equation}
Thomson scattering complicates the situation by the explicit angle coupling but
the solution can be obtained with the standard Feautrier scheme. Assuming
complete frequency redistribution in spectral lines \cite[p.\,29]{Mi78}, no
explicit frequency coupling occurs so that the parallel solution for all
frequencies enables a very efficient vectorization on the computer.

The following boundary conditions are used for the transfer equation. At the
inner boundary where the optical depth is at maximum, $\tau=\tau_{\rm max}$,
we have
\begin{equation}
\left(\mu \frac{\partial u_{\nu\mu}}{\partial\tau_{\nu}}\right)_{\tau_{\rm max}}
=I^+_{\nu\mu}-u_{\nu\mu}(\tau_{\rm max})
\end{equation}
where we specify $I^+_{\nu\mu}=I_{\nu}(+\mu, \tau_{\rm max})$ from the
diffusion approximation:
\begin{equation}\label{inner}
I^+_{\nu\mu}=B_{\nu}+\frac{3\mu}{\chi_{\nu}} \frac{\partial B_{\nu}}{\partial T} 
\frac{{\cal H}}
{\int_{0}^{\infty}\frac{1}{\chi_\nu}\frac{\partial B_{\nu}}{\partial T}\,d\nu}.
\end{equation}
$B_{\nu}$ is the Planck function and ${\cal H}$ the nominal (frequency
integrated) Eddington flux:
\begin{equation}\label{nominal}
{\cal H}=\sigma_R T_{\rm eff}^4/4\pi 
\end{equation}
with the Stefan-Boltzmann constant $\sigma_R$. At the outer boundary we take
$\tau_\nu=\tau_{\rm min}=m_1 \chi_\nu/2\rho$, assuming that $\chi$ is a linear
function of $m$ for $m<m_1$. Since $\tau_{\rm min}\neq 0$, it is not exactly
valid to assume no incident radiation at the stellar surface. Instead we
specify $I^-_{\nu\mu}=I_{\nu}(-\mu, \tau_{\rm min})$ after Scharmer \& Nordlund
\cite{SN82}:
\begin{equation}
I^-_{\nu\mu}=S_\nu(\tau_{\rm min})[1-\exp(-\tau_{\rm min}/\mu)]
\end{equation}
which follows from Eq.\,\ref{te} assuming $S(\tau)=S(\tau_{\rm min})$ for
$\tau<\tau_{\rm min}$. Then we get
\begin{equation}
\left(\mu \frac{\partial u_{\nu\mu}}{\partial\tau_{\nu}}\right)_{\tau_{\rm min}}
=u_{\nu\mu}(\tau_{\rm min})-I^-_{\nu\mu}.
\end{equation}
The boundary conditions are discretized performing Taylor expansions which 
yield second-order accuracy \cite[p.\,155]{Mi78}.

\subsection{Statistical equilibrium}

The statistical equilibrium equations are set up according to
\cite[p.\,127]{Mi78}. The number of atomic levels, ionization stages and
chemical species, as well as all radiative and collisional transitions are
taken from the input model atom supplied by the user (Sect.\,\ref{atom}).
Ionization into excited states of the next ionization stage is allowed for.
Dielectronic recombination and autoionization processes can also be included in
the model atom.

\subsubsection{Rate equations}

As usual the atomic energy levels are ordered sequentially by increasing
excitation energy, starting with the lowest ionization stage. Then for each
atomic level $i$ of any ionization stage of any species the rate equation
describes the equilibrium of rates into and rates out of this level:
\begin{equation}\label{rates}
n_i\sum_{i\neq j}^{}P_{ij}-\sum_{j\neq i}^{}n_jP_{ji}=0 .
\end{equation}
The rate coefficients $P_{ij}$ have radiative and collisional components:
$P_{ij}=R_{ij}+C_{ij}$. Radiative upward and downward rates are respectively
given by:
\begin{equation}
R_{ij}=4\pi\int_{0}^{\infty} \frac{\sigma_{ij}(\nu)}{h\nu}J_{\nu}\,d\nu
\end{equation}
\begin{equation}
R_{ji}=\left(\frac{n_i}{n_j}\right)^{\star}4\pi\int_{0}^{\infty} 
\frac{\sigma_{ij}(\nu)}
{h\nu}\left(\frac{2h\nu^3}{c^2}+J_{\nu}\right)e^{-h\nu/kT}\,d\nu .
\end{equation}
Photon cross-sections are denoted by $\sigma_{ij}(\nu)$.
$({n_i}/{n_j})^{\star}$ is the Boltzmann LTE population ratio in the case of
line transitions: $g_i/g_j \exp (-h\nu_{ij}/kT)$, where the $g_{i,j}$ are the
statistical weights. The LTE population number of a particular level is defined
relative to the ground state of the next ion, so that in the case of
recombination from a ground state $n_1^+$ we have by definition
$({n_i}/{n_j})^{\star}=n_e\phi_i(T)$ with the Saha-Boltzmann factor
\begin{equation}
\phi_i(T)=2.07 \cdot 10^{-16}\frac{g_i}{g_1^+}T^{-3/2}e^{h\nu_i/kT} 
\end{equation}
where $h\nu_i$ is the ionization potential of the level $i$. Care must be taken
in the case of recombination from an excited level into the next low ion. Then
$(n_i/n_j)^\star=n_e\phi_i \cdot \phi_1^+/\phi_j$.

Dielectronic recombination is included following \cite{MH73}. Assuming now that
$j$ is a ground state of ion $k$, then the recombination rate into level $i$ of
ion $k-1$ via an autoionization level $c$ (with ionization potential $h\nu_c$,
having a negative value when lying above the ionization limit) is:
\begin{equation}
R_{ji}=\frac{8\pi^2e^2}{mc^3}n_e\phi_if_{ic}e^{h(\nu_c-\nu_i)/kT}
\nu_c^2\left(1+\frac{c^2}{2h\nu_c^3\bar{J}}\right) .
\end{equation}
The reverse process, the autoionization rate, is given by:
\begin{equation}
R_{ij}=\frac{4\pi^2e^2}{hmc}\frac{1}{\nu_c}f_{ic}\bar{J} .
\end{equation}
The oscillator strength for the stabilizing transition (i.e.\ transition
i$\rightarrow$c) is denoted by $f_{ic}$, and $\bar{J}$ is the mean intensity
averaged over the line profile. The program simply takes $J_{\nu}$ from the
continuum frequency point closest to the transition frequency, which is
reasonable because the autoionization line profiles are extremely broad. The
population of autoionization levels is assumed to be in LTE and therefore such
levels do not appear explicitly in the rate equations.

The computation of collisional rates is generally dependent on the specific ion
or even transition. Several options, covering the most important cases, may be
chosen by the user.

\subsubsection{Abundance definition equation}

The rate equation for the highest level of a given chemical species is
redundant. It is replaced by the abundance definition equation. This equation
simply relates the total population of all levels of a particular species to
the total population of all hydrogen levels. Summation over all levels usually
includes not only NLTE levels but also levels which are treated in LTE,
according to the specification in the model atom. Denoting the number of
ionization stages of species $k$ with $NION(k)$, the number of NLTE and LTE
levels per ion with $NL(l)$ and $LTE(l)$, respectively, we can write:
\begin{equation}
\sum_{l=1}^{NION(k)}\left[\sum_{i=1}^{NL(l)}n_{kli}+
\sum_{i=1}^{LTE(l)}n_{kli}^{\star}\right]
=y_k \left[\sum_{i=1}^{NL(H)}n_{i}+\sum_{i=1}^{LTE(H)}n_{i}^{\star}+n_p\right]
. 
\end{equation}
On the right hand side we sum up all hydrogen level populations including the
proton density $n_p$, and $y_k$ is the number abundance ratio of species $k$ 
relative to hydrogen.

\subsubsection{Charge conservation}

We close the system of statistical equilibrium equations by invoking
charge conservation. We denote the total number of chemical species with 
$NATOM$, the charge of ion $l$ with $q(l)$ (in units of the electron charge) 
and write:
\begin{equation}
\sum_{k=1}^{NATOM}\sum_{l=1}^{NION(k)}q(l)
\left[\sum_{i=1}^{NL(l)}n_{kli}+\sum_{i=1}^{LTE(l)}n_{kli}^{\star}\right]=n_e .
\end{equation}

\subsubsection{Complete statistical equilibrium equations}

We introduce a vector comprising the occupation numbers of all NLTE levels,
$\bf n$ $=(n_1, \ldots ,n_{NL})$. Then the statistical equilibrium equation is
written as:
\begin{equation}\label{amat}
\bf An=b .
\end{equation}
The gross structure of the rate matrix $\bf A$ is of block matrix form, because
transitions between levels occur within one ionization stage or to the ground
state of the next ion. The structure is complicated by ionizations into
excited levels and by the abundance definition and charge conservation
equations which give additional non-zero elements in the corresponding lines
of ${\bf A}$.

\subsection{Radiative equilibrium}

Radiative equilibrium denotes the fact that the energy transport is exclusively
performed by photons. It can be enforced by adjusting the temperature
stratification either during the linearization procedure or in between ALI
iterations. In the former case a linear combination of two different
formulations is used and in the latter case a classical temperature correction
procedure (Uns\"old-Lucy), generalized to NLTE problems, is utilized. The
latter is particularly interesting, because it allows to exploit the blocked
form of the rate coefficient matrix. This will enable an economic block-by-block
solution followed by a subsequent Uns\"old-Lucy temperature correction step. On
the other side, however, this correction procedure may decelerate the global
convergence behavior of the ALI iteration.

\subsubsection{Differential and integral forms for linearization procedure}

The two forms of writing down the radiative equilibrium condition follow from
the postulation that the energy emitted by a volume element per unit time is
equal to the absorbed energy per unit time (integral form):
\begin{equation}\label{int}
\int_{0}^{\infty}\chi_{\nu}(S_{\nu}-J_{\nu})\,d\nu=0 ,
\end{equation}
where scattering terms in $\chi$ and $S_\nu$ cancel out. This formulation is
equivalent to invoking flux constancy throughout the atmosphere (differential
form) involving the nominal flux ${\cal H}$ (Eq.\,\ref{nominal}):
\begin{equation}\label{diff}
\int_{0}^{\infty}\frac{\partial}{\partial\tau_{\nu}}(f_{\nu}J_{\nu})\,d\nu-{\cal H}=0 
\end{equation}
where $f_{\nu}$ is the variable Eddington factor, defined as
\begin{equation}\label{eddfac}
f_{\nu}=\int_{0}^{1}\mu^2u_{\nu\mu}\,d\mu \Big/ \int_{0}^{1}u_{\nu\mu}\,d\mu
\end{equation}
and computed from the Feautrier variable $u_{\nu\mu}$ (Eq.\,\ref{udef}) after
the formal solution. As discussed e.g.\ in \cite{Hub88} the differential form
is more accurate at large depths, while the integral form behaves numerically
better at small depths. Instead of arbitrarily selecting that depth in the
atmosphere where we switch from one formulation to the other, we use a linear
combination of both constraint equations which guarantees a smooth transition
with depth, based on physical grounds \cite{HL92,Ham94}. Before adding up both
equations we have to take two measures. At first we divide Eq.\,\ref{int} by
the absorption mean of the opacity, ${\bar\kappa_J}$, for scaling reasons:
\begin{equation}\label{mean}
{\bar\kappa_J}=\frac{1}{J}\int_{0}^{\infty}\kappa_{\nu}J_{\nu}\,d\nu 
\qquad {\rm with} \quad J=\int_{0}^{\infty}J_{\nu}\,d\nu\, ,
\end{equation}
where $\kappa_\nu=\chi_\nu-n_e\sigma_e$ is the true opacity without electron
scattering. 
Then we multiply Eq.\,\ref{diff} with a similar average of the diagonal
elements of the $\Lambda^{\star}$ matrix:
\begin{equation}
{\bar\Lambda_J^\star}=\frac{1}{J}\int_{0}^{\infty}\Lambda^{\star}J_{\nu}\,d\nu\,.
\end{equation}
These two steps determine the relative weight of both equations in a particular
depth. Numerical experience shows that it is necessary to damp overcorrections
by adding the following term, which is computed from quantities of the previous
iteration step and which vanishes in the limit of convergence, to the right
hand side of Eq.\,\ref{diff}:
\begin{equation}
F_0\equiv  \left.\left({\bar\Lambda_J^\star}
\int_{0}^{\infty}\frac{\partial}{\partial\tau_{\nu}}(f_{\nu}J_{\nu})\,d\nu-{\bar\Lambda_J^\star}{\cal H}
\right)(1-{\bar\Lambda_J^\star})  \right|\raisebox{-.5cm}{\rm last iterate} .
\end{equation}
We write the equation of radiative equilibrium in its final form:
\begin{equation}\label{combi}
\frac{1}{\bar\kappa_J}\int_{0}^{\infty}\chi_{\nu}(S_{\nu}-J_{\nu})\,d\nu+
{\bar\Lambda_J^\star}\int_{0}^{\infty}\frac{\partial}{\partial\tau_{\nu}}(f_{\nu}J_{\nu})\,d\nu-
{\bar\Lambda_J^\star}{\cal H}-F_0=0 .
\end{equation}
We note that explicit depth coupling is introduced by the differential form
Eq.\,\ref{diff} through the derivative $\partial/\partial\tau_{\nu}$ even if a purely local
$\Lambda^{\star}$ operator is used. Therefore the linearization procedure can
no longer be performed independently at each depth point and the question
becomes relevant at which boundary to start with. Numerical experience shows
that it is essential to start at the outer boundary and to continue going
inwards. If a tri-diagonal operator is used, nearest neighbor depth coupling is
introduced anyhow. The program user can choose either the linear combination
Eq.\,\ref{combi} or the purely integral form Eq.\,\ref{int}, the latter may be
necessary to start the iteration under certain circumstances. The linear
combination, however, is found to give a much faster convergence behavior.

\subsubsection{Uns\"old-Lucy temperature correction procedure}\label{defm}

Closely following Lucy \cite{Lucy64} (but avoiding the Eddington approximation
and using variable Eddington factors instead) and generalizing to NLTE one can
derive for each depth point a temperature correction $\Delta T$ to be applied
to the actual temperature in order to achieve flux constancy. Using
$dx=-d\tau_\nu/\chi_\nu$, the zeroth momentum (i.e.\ angle averaged form) of the
radiation transfer Eq.\,\ref{te} is:
\begin{equation}\label{zmRT}
\frac{dH_\nu}{dx}=\chi_\nu\left(S_\nu-J_\nu\right).
\end{equation}
with $J_\nu$ from Eq.\,\ref{ali} and the Eddington flux $H_\nu$. In the LTE
case with electron scattering, $S_\nu$ can be written as the sum of a thermal
and a scattering contribution:
\begin{equation}
S_\nu = \frac{\kappa_\nu}{\chi_\nu}B_\nu +
\frac{n_e\sigma_e}{\chi_\nu}J_\nu.
\end{equation}
In the NLTE case we formally write in analogy:
\begin{equation}\label{gam}
S_\nu=\frac{\kappa_\nu^B}{\chi_\nu}B_\nu+\frac{\gamma_\nu}{\chi_\nu}J_\nu
\end{equation}
with quantities $\kappa_\nu^B$ and $\gamma_\nu$ which can be freely evaluated
but which are not independent of each other, since $\gamma_\nu$ must be
expressed by
\begin{equation}\label{gamma}
\gamma_\nu = \frac{\chi_\nu S_\nu-\kappa_\nu^B B_\nu}{J_\nu}
\end{equation}
in order to yield $S_\nu$ on the r.h.s.\ of Eq.\,\ref{gamma}. With this
substitution Eq.\,\ref{zmRT} reads:
\begin{equation}
\frac{dH_\nu}{dx}=\kappa_\nu^B B_\nu - \left(\chi_\nu -\gamma_\nu\right)J_\nu.
\end{equation}
Integrating over frequencies, the condition of flux conservation then reads:
\begin{equation}\label{zmRTi}
\frac{d{\cal H}}{d\tau} = \frac{\bar\kappa}{\bar\kappa_P}J-B\stackrel{!}{=}0
\end{equation}
where we used the following definitions for $\bar\kappa$, $\bar\kappa_P$, and $d\tau$:
\begin{equation}
\bar\kappa=\frac{1}{J}\int_{0}^{\infty}(\chi_\nu-\gamma_\nu )J_\nu\,d\nu=
\frac{1}{J}\int_{0}^{\infty}(\kappa_\nu(J_\nu-S_\nu)+\kappa_\nu^BB_\nu)d\nu,
\end{equation}
\begin{equation}
\bar\kappa_P=\frac{1}{B}\int_{0}^{\infty}\kappa_\nu^B B_\nu\,d\nu
\qquad {\rm and}
\qquad d\tau = -\bar\kappa_Pdx.
\end{equation}
Since we can choose $\kappa_\nu^B$ freely, we can define which opacities shall
contribute, finally resulting in a favorable scaling of factors in
Eq.\,\ref{lucy}. Usually we start with all processes included in $\kappa_\nu^B$
to begin with moderate corrections. Following Hauschildt (priv. comm.) one can
optionally exclude bound-bound or bound-free transitions which is necessary
if strong lines or continua dominate numerically the radiative equilibrium in
optically thin regions. Note that this measure does not affect the solution in
the case of convergence, but only the convergence rate. Without such an
acceleration, the Uns\"old-Lucy procedure may run into pseudo-convergence.

Integrating the first momentum of the radiation transfer equation over
frequency we obtain:
\begin{equation}\label{fmRTi}
\frac{dK}{d\tau}=\frac{\bar\kappa_H}{\bar\kappa_P}{\cal H}\qquad{\rm with}\qquad
\bar\kappa_H = \frac{1}{\cal H}\int_{0}^{\infty}\kappa_\nu H_\nu\,d\nu .
\end{equation}
Using Eq.\,\ref{zmRTi} and the depth integrated form of Eq.\,\ref{fmRTi} we
proceed as described by Lucy \cite{Lucy64}. We finally obtain, with frequency
averaged Eddington factors $\bar f$ and $\bar h$ as well as $\bar\kappa_S$
defined in analogy to Eq.\,\ref{mean}, the temperature correction at any depth:
\begin{equation}\label{lucy}
\hspace{-1cm}\Delta T=\frac{\pi}{4\sigma_R T^3} \frac{1}{\bar\kappa_P}
\left[ 
{\bar\kappa_J}J -{\bar\kappa_S}S + \frac{\bar\kappa}{\bar f}
\left(
\int_{0}^{m}\frac{\bar\kappa_H}{\rho}\Delta {\cal H}\,dm' + \frac{\Delta
  {{\cal H}(0)}{\bar f(0)}}{{\bar h(0)}}
\right)
\right]
\end{equation}

where $\Delta {\cal H}$ is the difference between the actual and the nominal
Eddington flux. In practice it is useful to accelerate this procedure by
extrapolating the last, say, ten corrections.

The Uns\"old-Lucy procedure provides model atmospheres with a relative
deviation from the flux constancy smaller than  10$^{-5}$ which is a factor of
ten better when compared to the procedure employing Eq.\,\ref{diff}. Due to the
decoupling of the temperature from the statistical equilibrium the
Uns\"old-Lucy  procedure is numerically much more stable allowing to calculate
models which otherwise failed to converge. The price is a slower overall
convergence of the ALI iteration by a factor of two.

\subsection{Hydrostatic equilibrium}

We write the equation for hydrostatic equilibrium as \cite[p.\,170]{Mi78}:
\begin{equation}
\frac{d}{dm}P=g
\end{equation}
where $g$ is the surface gravity and $m$ the column mass. $P$ is the total
pressure comprising gas, radiation and turbulent pressures, so that:
\begin{equation}\label{hydros}
\frac{d}{dm}\left( NkT+\frac{4\pi}{c}\int_{0}^{\infty}f_{\nu}J_{\nu}\,d\nu
+\frac{1}{2}\rho v^2_{\rm turb}\right)=g
\end{equation}
with Boltzmann's constant $k$ and the turbulent velocity $v_{\rm turb}$. The
hydrostatic equation may either be solved simultaneously with all other
equations or separately in between iterations. The overall convergence behavior
is usually the same in both cases. If taken into the linearization scheme and a
local $\Lambda^{\star}$ operator is used then, like in the case of the
radiative equilibrium equation, explicit depth coupling enters via the depth
derivative $d/dm$. Again, solution of the linearized equations has to proceed
inwards starting at the outer boundary. The starting value in the first depth
point (subscript $d=1$) is
\begin{equation}\label{eddfach}
N_1kT_1+\frac{1}{2}\rho_1v^2_{\rm turb}(m_1)=m_1\left(g-
\frac{4\pi}{c}\int_{0}^{\infty}
\frac{\chi_{1,\nu}}{\rho_1}h_{\nu}J_{\nu,k}\,d\nu\right)
\end{equation}
where $h_{\nu}$ is the variable Eddington factor denoting the ratio of
$H_{\nu}/J_{\nu}$ at the surface, kept fixed during linearization.

\subsection{Particle conservation}

The total particle density $N$ is the sum of electron density plus the
population density of all atomic states, LTE and NLTE levels. We may write down
the particle conservation equation in the following form that contains
explicitly only the hydrogen population numbers:
\begin{equation}
N=n_e+
\left[\sum_{i=1}^{NL(H)}n_{i}+\sum_{i=1}^{LTE(H)}n_{i}^{\star}+n_p\right]
\sum_{k=1}^{NATOM}y_k .
\end{equation}

\subsection{Fictitious massive particle density}

A fictitious massive particle density $n_H$ is introduced for notational
convenience. It is defined by
\begin{equation}
n_H=(N-N_e)\sum_{k=1}^{NATOM}m_ky_k \Big/ \sum_{k=1}^{NATOM}y_k .
\end{equation}
The mass of a chemical species in AMU is denoted by $m_k$. Introducing the mass
of a hydrogen atom $m_H$, we may simply write for the material density
\begin{equation}
\rho=n_Hm_H .
\end{equation}

\subsection{Opacity and emissivity \label{opa}}

Thermal opacity and emissivity are made up by atomic radiative bound-bound, bound-free
and free-free transitions. For each chemical species we compute and sum up:
\begin{eqnarray}\label{chi}
\kappa_{\nu}&=&\sum_{l=1}^{NION}\left[
\sum_{i=1}^{NL(l)}\sum_{j>i}^{NL(l)}\sigma_{li\rightarrow lj}(\nu)
\left(n_{li}-n_{lj}\frac{g_{li}}{g_{lj}}
e^{-h(\nu-\nu_{ij})/kT)}
\right) \right. \\ 
& & + \sum_{i=1}^{NL(l)}\sum_{j>i}^{NL(l+1)}\sigma_{li\rightarrow l+1,k}(\nu)
\left(n_{li}-n_{li}^{\star}e^{-h\nu/kT}\right) \nonumber \\ 
& & + \left. n_e\sigma_{kk}(l,\nu)\left(1-e^{-h\nu/kT}\right)
\left(\sum_{i=1}^{NL(l+1)}n_{l+1,i}+\sum_{i=1}^{LTE(l+1)}n_{l+1,i}^{\star}
\right)
\right] \nonumber
\end{eqnarray}
where the total opacity includes Thomson scattering, i.e.\
$\chi_\nu=\kappa_\nu+n_e\sigma_e$, and
\begin{eqnarray}\label{eta}
\frac{\eta_{\nu}}{2h\nu^3/c^2}&=&\sum_{l=1}^{NION}\left[
\sum_{i=1}^{NL(l)}\sum_{j>i}^{NL(l)}\sigma_{li\rightarrow lj}(\nu)
n_{lj}\frac{g_{li}}{g_{lj}}
e^{-h(\nu-\nu_{ij})/kT)} 
\right. \\
& & +\sum_{i=1}^{NL(l)}\sum_{j>i}^{NL(l+1)}\sigma_{li\rightarrow l+1,k}(\nu)
n_{li}^{\star}e^{-h\nu/kT} \nonumber \\
& & + \left. 
n_e\sigma_{kk}(l,\nu)e^{-h\nu/kT}
\left(\sum_{i=1}^{NL(l+1)}n_{l+1,i}+\sum_{i=1}^{LTE(l+1)}n_{l+1,i}^{\star}
\right)
\right] . \nonumber
\end{eqnarray}
The first index of variables marked with two indices denotes the ionization
stage and the second one denotes the ionic level. Thus $\sigma_{li\rightarrow
l+1,k}(\nu)$ denotes the cross-section for photoionization from level $i$ of
ion $l$ into level $k$ of ion $l+1$. The double summation over the bound-free
continua takes into account the possibility that a particular level may be
ionized into more than one level of the next high ion. Again, note the
definition of the LTE population number $n_{li}^{\star}$ in this case, which
depends on the level $(l+1,k)$ of the parent ion:
\begin{equation}
n_{li}^{\star}=n_{l+1,k}n_e\phi_{li}\frac{\phi_{l+1,1}}{\phi_{l+1,k}} .
\end{equation}
Note also, that the concept of LTE levels (whose population densities do enter,
e.g.\ the number or charge conservation equations) in the atomic models of
complex ions is therefore not unambiguous. The present code always assumes that
LTE levels in the model atoms are populated in LTE with respect to the ground
state of the upper ion.

The source function used for the approximate radiation transfer is the ratio
$\eta_{\nu}/\kappa_{\nu}$, thus, excludes Thomson scattering. For the exact
formal solution of course, the total opacity $\chi_\nu$ in the expression
Eq.\,\ref{source} includes the Thomson term ($n_e\sigma_e$).

\subsection{Atomic level dissolution by plasma perturbations}

\begin{figure}[bth]
\epsfxsize=\textwidth
\centerline{\epsffile{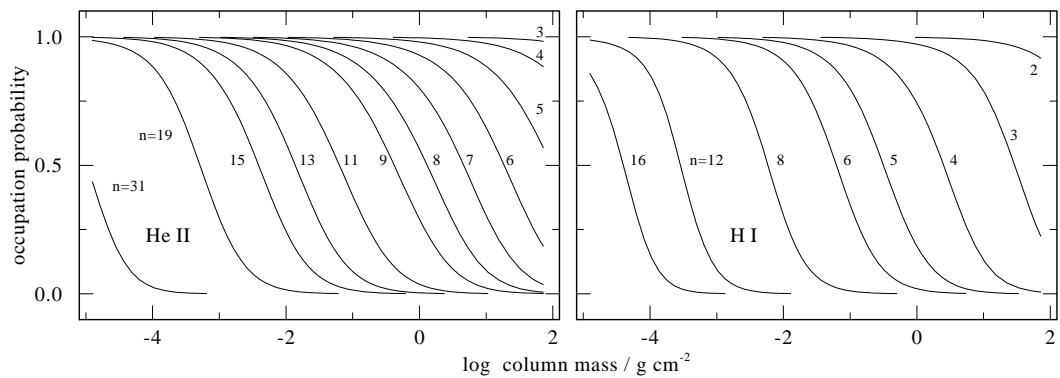}}
\caption{\small
Occupation probabilities of energy levels of ionized helium (left panel) and
hydrogen (right panel) as a function of depth in a white dwarf atmosphere.
Levels with high principal quantum number $n$ are already dissolved in the upper
atmosphere (left boundary of panels). At the inner boundary even all low-lying
levels are essentially dissolved. Atmospheric parameters are \Teff=100\,000\,K,
\logg=7.5, and H/He=0.1\%.
}
\label{humi}
\end{figure}

As high-lying atomic levels are strongly perturbed by other charged particles
in the plasma they are broadened and finally dissolved. This effect is
observable by line merging at series limits and has to be accounted for in line
profile analyses. Moreover, line overlap couples the radiation field in many
lines and flux blocking can strongly affect the global atmospheric structure.
Numerically, we treat the level dissolution in terms of occupation
probabilities, which for LTE plasmas can be defined as the ratio of the level
populations to those in absence of perturbations. A phenomenological theory for
these quantities was given in \cite{HuMi88}. The non-trivial generalization to
NLTE plasmas was performed by Hubeny \etal \cite{HHL94}. In practice an
individual occupation probability factor (depending on $T, n_e$, and principal
quantum number), is applied to each atomic level which describes the
probability that the level is dissolved. Furthermore, the rate equations
Eq.\,\ref{rates} must be generalized in a unique and unambiguous manner. For
details see \cite{HHL94}. As an example, Fig.\,\ref{humi} shows these
occupation probabilities for hydrogen and helium levels as a function of depth
in a white dwarf atmosphere.

\section{The Accelerated Lambda Iteration (ALI)}

In all constraint equations described above the mean intensities $J_{\nu}$ are
substituted by the approximate radiation field Eq.\,\ref{ali} in order to
eliminate these variables from the solution vector Eq.\,\ref{psi1}. In
principle the approximate lambda operator may be of arbitrary form as long as
the iteration procedure converges. In practice however an optimum choice is
desired in order to achieve convergence with a minimum amount of iteration
steps. The history of the ALOs is interesting and was summarized in detail by
Hubeny \cite{Hub92}. Of utmost importance were two papers by Olson and
collaborators \cite{OAB86,OK87} who overcame the major drawback of early ALOs,
namely the occurrence of free parameters controlling the convergence process,
and who found the optimum choice of ALOs. Our model atmosphere program enables
the use of either a diagonal or a tri-diagonal ALO, both are set up following
\cite{OK87}.

\subsection{Diagonal (local) lambda operators}

In this case the mean intensity $J_d$ at a particular depth  $d$ in the current
iteration step is computed solely from the local source function $S_d$ and a
correction term $\Delta J_d$, the latter involving the source functions (of all
depths) from the previous iteration. Dropping the iteration count and
introducing indices denoting depth points we can rewrite Eq.\,\ref{ali}:
\begin{equation}
J_d=\Lambda^{\star}_{d,d}S_d+\Delta J_d.
\end{equation}
In the discrete form we now think of $\Lambda^{\star}$ as a matrix acting on a
vector whose elements comprise the source functions of all depths. Then
$\Lambda^{\star}_{d,d}$ is the diagonal element of the $\Lambda^{\star}$ matrix
corresponding to depth point $d$. Writing $\Lambda^{\star}_{d,d}\equiv B_d$
(for numerical computation see Eq.\,\ref{matrix} below) we have a purely local
expression for the mean intensity:
\begin{equation}
J_d=B_dS_d+\Delta J_d
\end{equation}

\subsection{Tridiagonal (non-local) lambda operators}

Much better convergence is obtained if the mean intensity is computed not only
from the local source function but also from the source function of the
neighboring depths points. Then the matrix representation of $\Lambda^{\star}$
is of tri-diagonal form and we may write
\begin{equation}\label{trij}
J_d=C_{d-1}S_{d-1}+B_dS_d+A_{d+1}S_{d+1}+\Delta J_d
\end{equation}
where $C_{d-1}$ and $A_{d+1}$ represent the upper and lower subdiagonal
elements of $\Lambda^{\star}$ and $S_{d\pm 1}$ the source functions at the
adjacent depths. In analogy the correction term becomes
\begin{equation}\label{correc}
\Delta J_d=\Lambda_{d,d'}S_{d'}-(C_{d-1}S_{d-1}+B_dS_d+A_{d+1}S_{d+1}).
\end{equation}
Again all quantities for the computation of $\Delta J_d$ are from the previous
iteration and the first term denotes the exact formal solution of the transfer
equation. We emphasize again that the actual source functions in
Eq.\,\ref{trij} are computed from the actual population densities and
temperature which are unknown. We therefore have a non-linear set of equations
which is solved by either a Newton-Raphson iteration or other techniques,
resulting in the solution of a tri-diagonal linear equation of the form
Eq.\,\ref{tri}.

As was shown in \cite{OAB86} the elements of the optimum $\Lambda^{\star}$
matrix are given by the corresponding elements of the exact $\Lambda$ matrix.
The diagonal and subdiagonal elements are computed from \cite{OK87}:
\begin{eqnarray}\label{matrix}
A_{d+1}&= &\int_{0}^{1}\left(
e^{-\Delta\tau_d}\frac{e^{-\Delta\tau_{d-1}}-1}{\Delta\tau_{d-1}}
-\frac{e^{-\Delta\tau_d}-1}{\Delta\tau_d} \right)\frac{d\mu}{2} \nonumber\\
1-B_{d}&=&\int_{0}^{1}\left(
\frac{1-e^{-\Delta\tau_{d-1}}}{\Delta\tau_{d-1}}
+\frac{1-e^{-\Delta\tau_d}}{\Delta\tau_d} \right)\frac{d\mu}{2} \nonumber\\
C_{d-1}&= &\int_{0}^{1}\left(
e^{-\Delta\tau_{d-1}}\frac{e^{-\Delta\tau_{d}}-1}{\Delta\tau_{d}}
-\frac{e^{-\Delta\tau_{d-1}}-1}{\Delta\tau_{d-1}}\right)\frac{d\mu}{2}     
\end{eqnarray}
with $\Delta\tau_{d-1}\equiv (\tau_d-\tau_{d-1})/\mu$. At large optical depths
with increasing $\Delta\tau$ steps (the depth grid is equidistant in $\log\tau$)
the subdiagonals $A_{d+1}$ and $C_{d-1}$ vanish and the diagonal $B_d$
approaches unity, resembling the fact that the radiation field is more and more
determined by local properties of the matter. At very small optical depths all
elements of $\Lambda^{\star}$ vanish, reflecting the non-localness of the
radiation field in this case.

\subsection{Acceleration of convergence}

{\tt PRO2} allows usage of an acceleration scheme to speed up convergence of
the iteration cycle Eq.\,\ref{ali}. We implemented the scheme originally
proposed by Ng \cite{Ng74,Auer87}. It extrapolates the correction vector
$\delta\mbox{\boldmath$\psi$}_d$ from the previous three iterations. From our
experience the extrapolation often yields over-corrections resulting in
alternating convergence or even divergence. And usually the application of a
tri-diagonal ALO results in a satisfactorily fast convergence so that the
acceleration scheme is rarely used.

\section{Solution of the non-linear equations by iteration}

The complete set of non-linear equations for a single iteration step
Eq.\,\ref{ali} comprises at each depth the equations for statistical,
radiative, and hydrostatic equilibrium and the particle conservation equation.
For the numerical solution we introduce discrete depth and frequency grids. The
equations are then linearized and solved by a suitable iterative scheme.
Explicit angle dependency of the radiation field is not required here and
consequently eliminated by the use of variable Eddington factors. Angle
dependency is only considered in the formal solution of the transfer equation.
The program requires an input model atmosphere structure as a starting
approximation together with an atomic data file, as well as a frequency grid.
Depth and frequency grids are therefore set up in advance by separate programs.

\subsection{Discretization} 

A depth grid is set up by an auxiliary program which computes, starting from a
gray approximation, a LTE continuum model using the Uns\"old-Lucy temperature
correction procedure. In this program depth points are set equidistantly on a
logarithmic (Rosseland) optical depth scale. The user may choose the inner and
outer boundary points and the total number of grid points (typically 90). The
converged LTE model (temperature and density structure, given on a column mass
depth scale) is written to a file that is read by {\tt PRO2}. The NLTE code
uses the column mass as an independent depth variable.

The frequency grid is established based upon the atomic data input file
(see Sect.\,\ref{atom}). Frequency points are set blue- and redward of each
absorption edge and for each spectral line. Gaps are filled up by setting
continuum points. Finally, the quadrature weights are computed. The user may
change default options for this procedure. Frequency integrals appearing e.g.\
in Eq.\,\ref{combi} are replaced by quadrature sums and differential quotients
involving depth derivatives by difference quotients.

\subsection{Linearization} 

All variables $x$ are replaced by $x\rightarrow x+\delta x$ where $\delta x$
denotes a small perturbation of $x$. Terms not linear in these perturbations
are neglected. The perturbations are expressed by perturbations of the basic
variables:
\begin{equation}\label{deltax}
\delta x=      \frac{\partial x}{\partial T  }\delta T  +
               \frac{\partial x}{\partial n_e}\delta n_e+
               \frac{\partial x}{\partial N  }\delta N  +
               \frac{\partial x}{\partial n_H}\delta n_H+
\sum_{l=1}^{NL}\frac{\partial x}{\partial n_l}\delta n_l .
\end{equation}
As an illustrative example we linearize the equation for radiative equilibrium.
Most other linearized equations may be found in \cite{W86}. Assigning two
indices ($d$ for depth and $i$ for frequency of a grid with NF points) to the
variables and denoting the quadrature weights with $w_i$ Eq.\,\ref{combi}
becomes:
\begin{eqnarray}\label{lin}
\lefteqn{
\sum_{i=1}^{NF}
w_i(\frac{\chi\di}{\bar\kappa_J}[\delta S\di-\delta J\di]+\delta\chi\di[S\di-J\di]) 
}\nonumber\\
& &
+{\bar\Lambda_J^\star}\sum_{i=1}^{NF} 
\frac{w_i}{\Delta\tau_i}(\delta J\di f\di-\delta J\dmi f\dmi) 
= F_0+{\bar\Lambda_J^\star}{\cal H}- \nonumber\\
& &
\sum_{i=1}^{NF}w_i\frac{\chi\di}{\bar\kappa_J}(S\di-J\di)
-{\bar\Lambda_J^\star}\sum_{i=1}^{NF}\frac{w_i}{\Delta\tau_i}(f\di J\di-f\dmi J\dmi) .
\end{eqnarray}
Note that we do not linearize $\Delta\tau_i$. Because of this, convergence
properties may be significantly deteriorated in some cases. Perturbations
$\delta S\di, \delta\chi\di$ are expressed by Eq.\,\ref{deltax}, and the
perturbation of the mean intensity $J\di$ is, according to Eq.\,\ref{trij},
given through the perturbations of the source function at the actual and the
two adjacent depths:
\begin{equation}
\delta J\di=C\dmi\delta S\dmi+B\di\delta S\di+A\dpi\delta S\dpi
\end{equation}
where $A, B, C$ are the $\Lambda$ matrix elements from Eq.\,\ref{matrix}. The
$\delta J\dmi$ involve the term $C\dmmi\delta S\dmmi$ which is neglected
because we only want to account for nearest neighbor coupling. We write $\delta
S\di$ with the help of Eq.\,\ref{deltax} and observe that for any variable $z$
\begin{equation}
\frac{\partial S\di}{\partial z}=\frac{1}{\chi\di}\left(
\frac{\partial\eta\di}{\partial z}-S\di\frac{\partial\chi\di}{\partial z}
\right).
\end{equation}
Derivatives of opacity and emissivity with respect to temperature, electron and
population densities are computed from analytical expressions (see e.g.\
\cite{MHA75,W87}). We finally get from Eq.\,\ref{lin}:
\begin{eqnarray}
 & & 
\delta T\dmi\left\{
\sum_{i}^{NF}-\frac{w_i}{\bar\kappa_J}
\frac{\partial S\dmi}{\partial T}\chi\di C\dmi \right.\nonumber
\\
 & &\left. +{\bar\Lambda_J^\star}\sum_{i}^{NF}\frac{w_i}{\Delta\tau_i}
(f\di C\dmi-f\dmi B\dmi)\frac{\partial S\dmi}{\partial T}
\right\}+
\nonumber
\\
 & & 
\delta T_d\left\{
\sum_{i}^{NF}\frac{w_i}{\bar\kappa_J}\left[
\frac{\partial S\di}{\partial T}\chi\di(1-B\di)+
\frac{\partial\chi\di}{\partial T}(S\di-J\di)\right] \right.\nonumber
\\
 & &\left.+{\bar\Lambda_J^\star}\sum_{i}^{NF}\frac{w_i}{\Delta\tau_i}(f\di B\di-f\dmi A\di)
\frac{\partial S\di}{\partial T}
\right\}+
\nonumber\\
 & & 
\delta T\dpi\left\{
\sum_{i}^{NF}-\frac{w_i}{\bar\kappa_J}
\frac{\partial S\dpi}{\partial T}\chi\di A\dpi \right.\nonumber
\\
& & \left. +{\bar\Lambda_J^\star}\sum_{i}^{NF}\frac{w_i}{\Delta\tau_i}
(f\di A\dpi-f\dmi B\dpi)\frac{\partial S\dpi}{\partial T}
\right\}+
\nonumber
\\
& & 
\delta n_{e_{d-1,i}}\{\cdots\}
    +\delta n_{e_{d,i}}\{\cdots\}
    +\delta n_{e_{d+1,i}}\{\cdots\}+
\nonumber
\\
& & 
\sum_{l=1}^{NL}\delta n_{l_{d-1,i}}\{\cdots\}
    +\sum_{l=1}^{NL}\delta n_{l_{d,i}}\{\cdots\}
    +\sum_{l=1}^{NL}\delta n_{l_{d+1,i}}\{\cdots\}
\nonumber
\\
 &  & 
= {\rm r.h.s.}
\end{eqnarray}
Curly brackets $\{\cdots\}$ denote terms that are similar to those
multiplied with the perturbations of the temperature. Instead of partial
derivatives in respect to $T$, they contain derivatives in respect to $n_e$ and
the populations $n_l$. They all represent coefficients of the matrices
{$\mbox{\boldmath$\alpha, \beta, \gamma$}$} in Eq.\,\ref{tri}.

\subsection{Newton-Raphson iteration}

As described in Sect.\,\ref{eins2} the linearized equations have a tri-diagonal
block-matrix form, see Eq.\,\ref{tri}. Inversion of the grand matrix ($\equiv
{\bf T}$ sized $(NN\cdot ND) \times (NN\cdot ND)$, i.e.\ about $10^4\times 10^4$
in typical applications) is performed with a block-Gaussian elimination scheme,
which means that our iteration of the non-linear equations represents a
multi-dimensional Newton-Raphson method. The problem is structurally simplified
when explicit depth coupling is avoided by the use of a local ALO, however, the
numerical effort is not much reduced, because in both cases the main effort
lies with the inversion of matrices sized $NN\times NN$. The Newton-Raphson
iteration involves two numerically expensive steps, first setting up the
Jacobian (comprising {$\mbox{\boldmath$\alpha, \beta, \gamma$}$}) and then
inverting it. Additionally, the matrix inversions in Eq.\,\ref{feau} limit
their size to about $NN=150$ because otherwise numerical accuracy is lost. Two
variants recently introduced in stellar atmosphere calculations are able to
improve both, numerical accuracy and, most of all, computational speed.

\subsection{Alternative fast solution techniques for non-linear equations: 
Broyden-- and Kantorovich--variants}

Broyden's variant \cite{Broy65} belongs to the family of so-called quasi-Newton
methods and it was first used in model atmosphere calculations in
\cite{DW91,HKW91,KHK92}. It avoids the repeated set-up of the Jacobian by the
use of an update formula. On top of this, it also gives an update formula for the
{\em inverse} Jacobian. In the case of a local ALO the solution of the
linearized system at any depth is
\begin{equation}\label{yyy}
\delta\mbox{\boldmath$\psi$} = \mbox{\boldmath$\beta$}^{-1}_k \, {\bf c} .
\end{equation}
Let $\mbox{\boldmath$\beta$}^{-1}_k$ be the $k$-th iterate of the inverse
Jacobian, then an update can be found from:
\begin{equation}\label{xxx}
\mbox{\boldmath$\beta$}_{k+1}^{-1}=\mbox{\boldmath$\beta$}_k^{-1}
+\frac{({\bf s}_k-\mbox{\boldmath$\beta$}_k^{-1}{\bf y}_k)
\otimes({\bf s}_k^T\mbox{\boldmath$\beta$}_k^{-1})}{{\bf s}_k^T
\mbox{\boldmath$\beta$}_k^{-1}{\bf y}_k}
\end{equation}
where $\otimes$ denotes the dyadic product and where we have defined:
\begin{eqnarray}
{\bf s}_k  \equiv & \delta\mbox{\boldmath$\psi$}_k 
& \quad \mbox{solution vector of preceding linearization} \nonumber\\
{\bf y}_k  \equiv & {\bf c}_{k+1}-{\bf c}_k 
& \quad \mbox{difference of actual and preceding residuum} . \nonumber 
\end{eqnarray}
The convergence rate is super-linear, i.e.\ slower than the quadratic
rate of the Newton-Raphson method, but this is more than compensated by the tremendous
speed-up for a single iteration step. It is not always necessary to begin the
iteration with the calculation of an exact Jacobian and its inversion.
Experience shows that in an advanced stage of the overall (ALI-) iteration
Eq.\,\ref{ali} (i.e.\ when corrections become small, of the order 1\%) we can
start the linearization cycle Eq.\,\ref{xxx} by using the inverse Jacobian
from the previous overall iteration. Computational speed-up is extreme in this
case, however, it requires storage of the Jacobians of all depths.

More difficult is the application to the tri-diagonal ALO case. Here we have
to update the grand matrix ${\bf T}$ which, as already mentioned, is of block tri-diagonal
form. We cannot update their inverse, because it is never computed explicitly.
Furthermore we need an update formula that preserves the block tri-diagonal
form which is a prerequisite for its inversion by the Feautrier scheme
Eq.\,\ref{feau}. Such a formula was found by Schubert \cite{Schu70}:
\begin{equation}
{\bf T}_{k+1}={\bf T}_k
+\frac{({\bf y}_k-{\bf T}_k{\bf s}_k)\otimes{\bf\bar s}_k^T}
{{\bf\bar s}_k^T{\bf\bar s}_k}
\end{equation}
where ${\bf\bar s}_k\equiv {\bf Z}{\bf s}_k$ with the structure matrix ${\bf Z}$
as defined by:
\[ Z_{ij}=\left\{ \begin{array}{r@{\quad {\rm if}\quad}l}
                   1 & T_{ij}\neq 0 \\ 0 & T_{ij}=0 . 
                  \end{array}  \right.\] 
The vectors ${\bf s}_k$ and ${\bf y}_k$ are defined as above but now they span
over the quantities of all instead of a single depth point. With this formula
we obtain new submatrices {$\mbox{\boldmath$\alpha, \beta, \gamma$}$} and ${\bf
c}$ with which the Feautrier scheme Eq.\,\ref{feau} is solved again. This
procedure saves the computation of derivatives. Another feature realized in our
program also saves the repeated inversion of ${\bf q}\equiv
(\mbox{\boldmath$\beta$}_d+\mbox{\boldmath$\gamma$}_d{\bf D}_{d-1})$ by
updating its inverse with the Broyden formula Eq.\,\ref{xxx}. Similar to the
diagonal ALO case it is also possible to pass starting matrices from one
overall iteration Eq.\,\ref{ali} to the next for the update of $\bf T$ and the
matrix ${\bf q}^{-1}$. In both cases the user specifies two threshold values
for the maximum relative correction in $\delta\mbox{\boldmath$\psi$}$ which
cause the program to switch from Newton-Raphson to Broyden stages 1 and 2.
During stage 1 each new overall cycle Eq.\,\ref{ali} is started with an exact
calculation and inversion of all matrices involved and in stage 2 these
matrices are passed through each iteration.

Another variant, the Kantorovich method was recently introduced into model
atmosphere calculations \cite{HL92}. It is more simple and straightforward to
implement. This method simply keeps fixed the Jacobian during the linearization
cycle and it is surprisingly stable. In fact it turns out to be even more
stable (i.e.\ it can be utilized in an earlier stage of iteration) than the
Broyden method in the tri-diagonal ALO case. The user of {\tt PRO2} may
choose this variant in two stages in analogy to the Broyden variant. It was
found that in the stage 2 it is necessary to update the Jacobian,
say, every 5 or 10 overall iterations in order to prevent divergence.

\section{NLTE metal line blanketing}                                     

Despite the capacity increase for the  NLTE treatment of model atmosphere
problems by introducing the ALI method combined with pre-conditioning
techniques, the blanketing by millions of lines from the iron group elements
arising from transitions between some $10^5$ levels could only be attacked with
the help of statistical methods. These have been introduced into NLTE model
atmosphere work by Anderson \cite{And89,And91}. At the outset, model
atoms are constructed by combining many thousand of levels into a relatively
small number of superlevels which can be treated with ALI (or other) methods.
Then, in order to reduce the computational effort, two approaches were
developed which vastly decrease the number of frequency points (and hence the
number of transfer equations to be solved) to describe properly the complex
frequency dependence of the opacity. These two approaches have their roots in
LTE modeling techniques, where for the same reason statistical methods are
applied for the opacity treatment: The Opacity Distribution Function (ODF) and
Opacity Sampling (OS) approaches. Both are based on the circumstance that the
opacity (in the LTE approximation) is a function of two only local
thermodynamic quantities. Roughly speaking, each opacity source can be written
in terms of a population density and a photon cross-section for the respective
radiative transition:\\ 
\centerline{$\kappa_\nu \sim n_l \sigma_{lu}(\nu)$}
\\ \\
In LTE the population follows from the Saha-Boltzmann equations, hence
$n_l=n_l(n_e,T)$. The OS and ODF methods use such pre-tabulated (on a very fine
frequency mesh) $\kappa_\nu(n_e,T)$ during the model atmosphere calculations. The
NLTE situation  is more complicated, because pre-tabulation of opacities is not
useful. The population densities at any depth now also depend explicitly on the
radiation field (via the rate equations which substitute the TE Saha-Boltzmann
statistics) and thus on the populations in each other depth of the atmosphere.
As a consequence, the OS and ODF methods are not applied to opacity
tabulations, but on tabulations of the photon cross-sections $\sigma(\nu)$.
These do depend on local quantities only, e.g.\ line broadening by Stark and
Doppler effects is calculated from $T$ and $n_e$. In the NLTE case the
cross-section takes over the role which the opacity played in the LTE case. So,
strictly speaking, the designation OS and ODF is not quite correct in the NLTE
context.

The strategy in our code is the following. Before any model atmosphere
calculation is started, the atomic data are prepared by constructing
superlevels,  and the cross-sections for superlines. Then these cross-sections
are either sampled  on a coarse frequency grid or ODFs are constructed. These
data are put into the model atom which is read by the code. The code does not
know if OS or ODFs are used, i.e.\ it is written to be independent of any of
these approaches.

\subsection{Model atoms for iron group elements}

\begin{figure}[bth]
\epsfxsize=9.1cm
\centerline{\epsffile{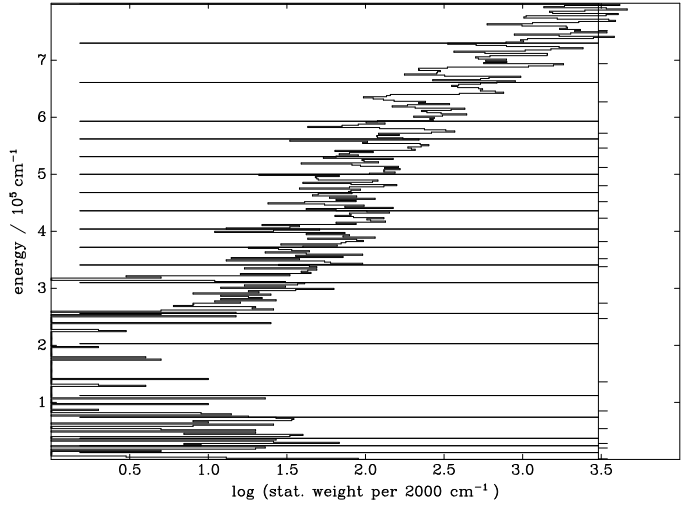}}
\caption{\small
Energy distribution of statistical weights of the iron group elements in
ionization stage VI. Individual energy levels are grouped into bands
(horizontal lines) and merged into superlevels with an average energy.
}
\label{levelfig}
\end{figure}

The large number of atomic levels in a single ionization stage is grouped into
a small number of typically 10--20 superlevels or, energy bands. Grouping is
performed by inspecting a level diagram (Fig.\,\ref{levelfig}) which shows the
number of levels (times their statistical weight) per energy bin as a function
of excitation energy. Gaps and peaks in this distribution are used to define
energy bands. Each of these bands is then treated as a single NLTE level with
suitably averaged statistical weight and energy. All individual lines
connecting levels out of two distinct bands are combined to a band-band
transition with a so-called complex photon cross-section. This cross-section
essentially is a sum of all individual line profiles which however conserves
the exact location of the lines in the frequency spectrum. This co-addition is
performed once and for all and on a very fine frequency mesh to account for the
profile shape of every line, before any model atmosphere calculation begins.
These complex cross-sections (examples are seen in the top panels of
Figs.\,\ref{xfig} and \ref{odffig}) are tabulated and later used to construct
ODFs or to perform OS for the model calculations.

Each of the model bands $L$ is treated as one single NLTE level with an
average energy $\EL{L}$ and statistical weight $\GL{L}$ which are computed
from all the individual levels ($\EL{l}$, $\gl{l}$) within a particular band:
\begin{equation}
\EL{L}=\sum_{l\epsilon L} \EL{l}\gl{l}^*\left/\sum_{l\epsilon L}\gl{l}^*\right. 
\qquad \GL{L}=e^{\EL{L}/kT^*}\sum_{l\epsilon L} \gl{l}^*\  
\end{equation}
where $\gl{l}^*=\ai\gl{l}e^{-\EL{l}/kT^*}$. 
$T^*$ is a characteristic temperature, pre-chosen and fixed throughout the
model calculations, and at which the ionization stage in question is most
strongly populated. Energy levels of all iron group elements $s$ in the same
ionization stage contribute to these model bands according to their abundance
$a_s$. All individual line transitions with cross-sections $\slu{lu}$ between
two model bands $L$ and $U$ are combined to one complex band-band transition
with a cross-section $\slu{LU}$ as described by:
\begin{equation}
\slu{LU}(\nu )=
\frac{\pi e^{2}}{m_ec}\frac{1}{ \sum_{l\epsilon L}\gl{l}^*}
\sum_{l\epsilon L,u\epsilon U}\gl{l}^{*}\flu{lu}\phi(\nulu{lu}-\nu )\,.
\end{equation}
$\phi(\nulu{lu}-\nu )$ is the normalized profile of an individual line. This
means that all individual lines are correctly accounted for in a sense that
their real position within the frequency spectrum is not affected by the
introduction of atomic model bands. The complex cross-sections (each possibly
representing many thousand individual lines) are computed in {\em advance} of
the model atmosphere calculations on a fine frequency grid with a resolution
smaller than one thermal Doppler width (typically $0.1$ $\Delta\nu_D$). This is
done at two values for the electron density ($0$ and $10^{16}\,{\rm cm}^{-3}$)
and the NLTE code accounts for depth dependent electron collisional broadening
by interpolation. Individual line photon cross-sections are represented by
Voigt profiles including Stark broadening.

Collisional excitation rates between atomic model bands are treated with a
generalized Van Regemorter \cite{Vanreg62} formula:
\begin{equation}
\Clu=\pi a_{0}^2 \left(\frac{8k}{m_e\pi}\right)^{1/2}T^{1/2}
n_{e}e^{-\Elu/kT}\Glu(T^*) 
\end{equation}
\begin{equation}
\hspace{-.5cm}{\rm with} \qquad \Glu (T^*)=
\frac{14 E_H^2}{kT^*}\frac{1}{\sum_{l\epsilon L}\gl{l}^*}
\sum_{lu}\gl{l}^{*}\flu{lu}\frac{P(E_{lu}/kT^*)}{\EL{lu}}e^{(\Elu-\EL{lu})/kT^*}
\end{equation}
where $P(x)={\rm max}[\bar{g},0.276 e^x E_1(x)]$. $E_H$ is the ionization
potential of hydrogen (in electron volts), $E_1$ is the first exponential
integral and $\bar{g}$ a constant depending on the ionic charge. The $\Glu$
involve the f-values of all individual lines and they are computed together
with the radiative cross-sections. Third degree polynomials in $\log kT^*$ are
fitted to $\Glu$ and the coefficients are written into the atomic input data
file for the NLTE code.

Photoionization cross-sections for iron group elements have numerous strong
resonances that are difficult to deal with. As a first approximation one can
calculate hydrogen-like cross-sections $\slu{l,bf}$ for the individual levels
$l$ and combine them to a complex ionization cross-section for every model band:
\begin{equation}
\slu{BF}(\nu)=e^{-\EL{L}/kT^*}\sum_{l\epsilon L}\gl{l}^{*}\slu{l,bf}(\nu).
\end{equation}
This cross-section is stored in a file and read by the code. Other data to be
used alternatively (available e.g.\ from the Opacity Project) may easily be
prepared and stored in such a file by the user. For collisional ionization one
may select Seaton's \cite{Sea62} formula with a mean (hydrogen-like) ionization
cross-section.

\subsection{OS and ODF approaches}

The OS or, alternatively, ODF approaches are introduced merely in order to save
computing time during the model atmosphere calculations. In principle it is
possible to proceed directly with the complex cross-sections constructed as
described above. However, this would require a very fine frequency mesh over
the entire spectrum in order to discretize the cross-sections in a similar
detailed manner, resulting in some $10^5$ frequency points. Since computation
time scales linearly with the number of frequency points in the ALI formalism,
a reduction to some thousand or ten thousand points easily reduces the
computational effort by an order of magnitude.

\begin{figure}[bth]
\epsfxsize=9.1cm
\centerline{\epsffile{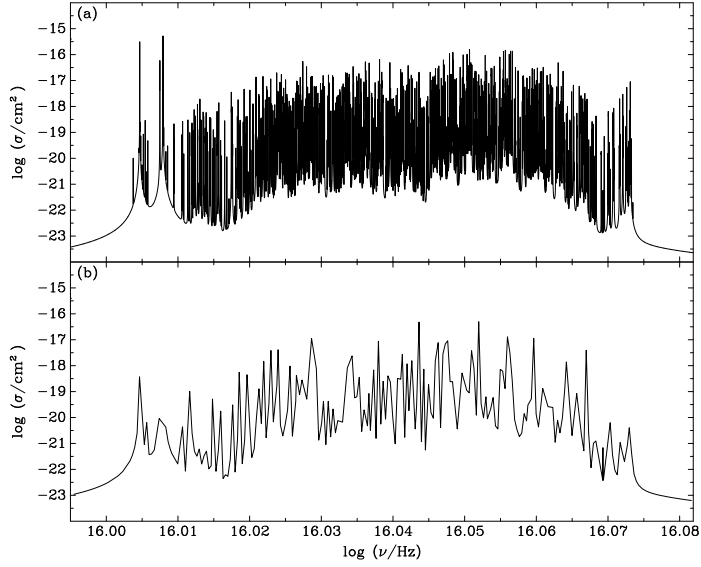}}
\caption{\small
More than 1000 line transitions between individual levels from two specific
superlevels (Fig.\,\ref{levelfig}) are co-added to a complex photon
cross-section resolved by 330\,000 frequency points (top panel). Sampling of
this cross section with 300 points results in the cross section shown in the
bottom panel.
}
\label{xfig}
\end{figure}

Opacity Sampling is the more straightforward approach. The fine cross-section
is sampled by a coarse frequency grid and the resulting coarse cross-section is
used for the model calculation (Fig.\,\ref{xfig}). Individual lines are no
longer accounted for in an exact way, but this is not necessary in order to
account for the line blanketing effects, i.e.\ effects of metal lines on the
global atmospheric structure like surface cooling and backwarming of deeper
layers. A high resolution synthetic spectrum can be obtained easily after model
construction by performing a single solution of the radiation transfer equation
on a fine frequency mesh.

The quality of the sampling procedure can be checked by a quadrature of the
cross-section on the frequency grid (with weights $w_k$):
\begin{equation}
\frac{\pi e^2}{m_ec}\frac{1}{\sum_{l\epsilon L}\gl{l}^*}
\sum_{lu}\gl{l}^*\flu{lu}\stackrel{!}{=} \sum_{k}\slu{LU}(k)w_k\,.
\end{equation}
Renormalization may be performed if necessary. This reduction of the
cross-sections by sampling is also performed before the model calculations
begin.

\begin{figure}[bth]
\epsfxsize=9.1cm
\centerline{\epsffile{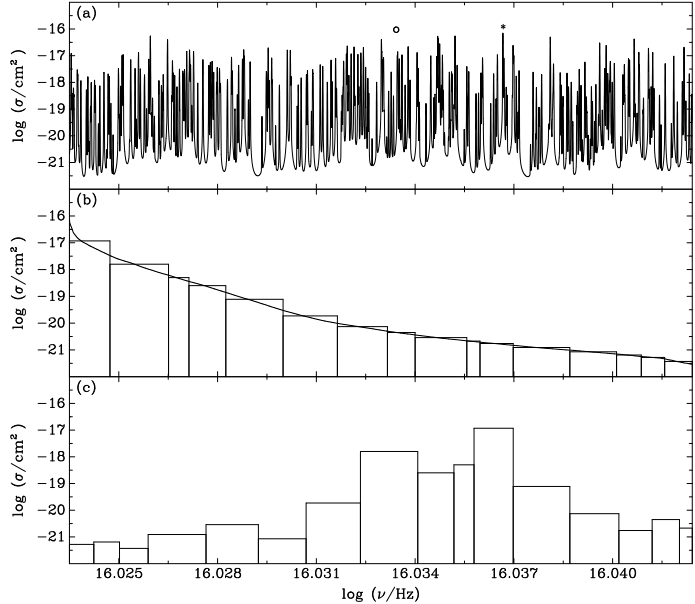}}
\caption{\small
Construction of a cross-section distribution function and its representation by
a step function (middle panel) from a portion of a complex cross-section (top).
In the bottom panel a randomized arrangement of the interval steps is shown.
}
\label{odffig}
\end{figure}

The alternative way is the construction of Opacity Distribution Functions  (or,
more correctly, cross-section distribution functions). Each complex
cross-section is re-ordered in such a way that the resulting ODF is a monotonous
function (see Fig.\,\ref{odffig}, middle panel). The resulting smooth run of
the cross-section over frequency can be approximated by a simple step function
with typically one dozen of intervals. This cross-section is then fed into the
model atmosphere code which can use a coarse frequency mesh to appropriately
incorporate the ODFs. In order to avoid unrealistic systematic effects,
however, the cross-section bins within each ODF are re-shuffled randomly
(bottom panel of Fig.\,\ref{odffig}).

Many numerical tests concerning model atom construction with superlevels were
performed by studying the effects of details in band definition and widths.
Also, the resulting model atmospheres using ODF and OS approaches were compared
and generally, good agreement was found \cite{Haas97}.

\subsection{Atomic data and model atoms}\label{atom}

All recent progress in stellar atmosphere modeling would have been impossible
without the availability of atomic input data. Major data sources were put at
public disposal by the Opacity Project \cite{Sea94} and the work by Kurucz
\cite{Kur91}. These sources provide energy levels, transition probabilities,
and bound-free photon cross-sections. The Iron Project \cite{Hum93} is
delivering electron collision strengths which are important for NLTE
calculations and which were hardly available up to now. We cannot
over-emphasize these vital contributions to our work.

The atomic species that are to be included in the model atmosphere calculations
are entirely determined by an atomic data input file. For each ionization stage
of any chemical element the user defines atomic levels by the ionization
potential and statistical weight and assigning a name (character string) to
them. These level names are used to define radiative and collisional
bound-bound and bound-free transitions among the levels as well as free-free
opacities. The declaration of such transitions is generally complemented by a
number which specifies the formula which {\tt PRO2} shall use to calculate
cross-sections for the rates and opacities. Depending on the formula chosen by
the user, additional input data are occasionally expected, such like oscillator
strengths or line broadening data. Alternatively, photon cross-sections for
lines and continua may be read from external files whose names need to be
declared with the definition of the transitions. Construction of model atoms
involving large datasets, e.g.\ from the Opacity Project, is automated and
requires a minimum of work by the user. The interested reader is referred to a
comprehensive User's Guide for {\tt PRO2} available from the authors \cite{WRD98}.

\section{Application to hot compact stars}

One important motivation for developing and applying the ALI method for stellar
atmospheres was the unsolved problem of NLTE metal line blanketing in hot
stars. We want to focus here on two topics which highlight the successful
application of the new models. The first concerns the Balmer line problem which
until recently appeared to be a fundamental drawback of NLTE models. The second
example describes the abundance determination of iron group elements in evolved
compact stars by constructing self-consistent models which can reproduce
simultaneously the observed spectral properties of white dwarfs and subdwarf O
stars (sdO) from the optical region through the extreme ultraviolet regime.

\subsection{Balmer line profiles}

\begin{figure}[bth]
\epsfxsize=11cm
\centerline{\epsffile{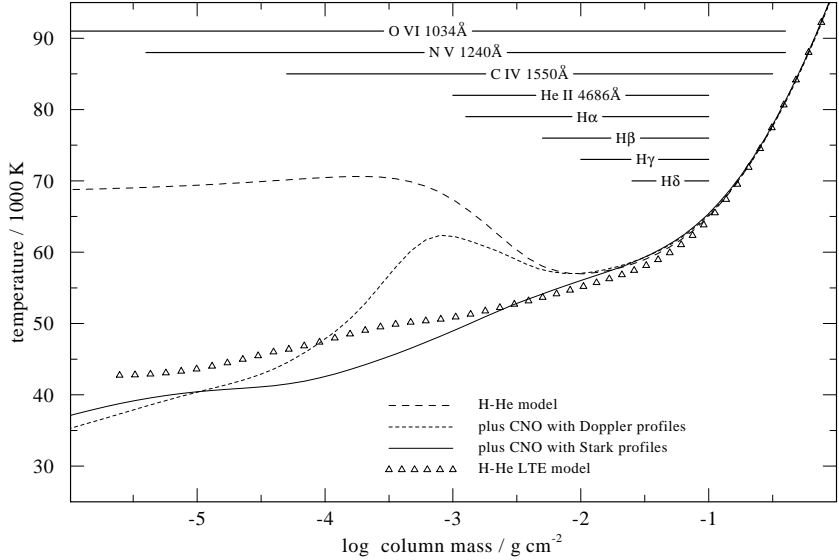}}
\caption{\small
Temperature structures of three NLTE model atmospheres with increasing degree
of sophistication. A LTE model is shown for comparison (triangles). While pure
H-He models show a high-temperature plateau at the surface, our most
sophisticated model shows a monotonous temperature run (full line). Formation
depth intervals for selected lines are represented by horizontal bars.
\Teff=82\,000\,K, \logg=6.2, solar abundance ratios.
}
\label{tempfig}
\end{figure}

\begin{figure}[bth]
\epsfxsize=10cm
\centerline{\epsffile{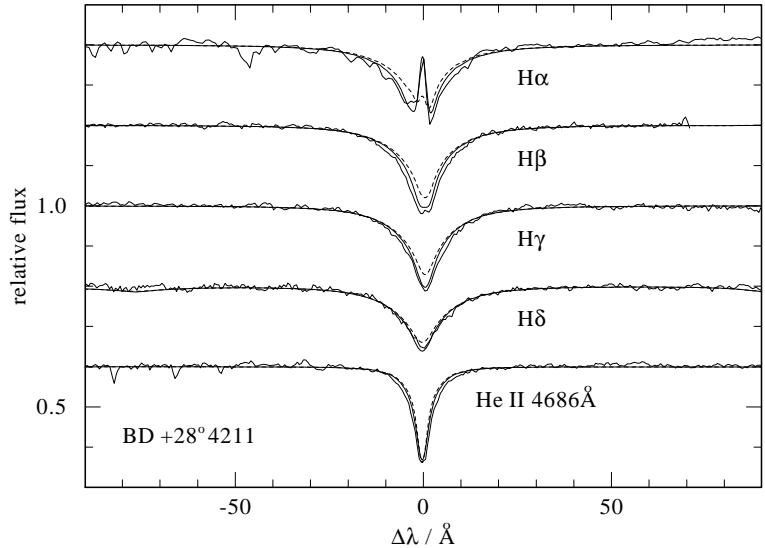}}
\caption{\small
Line profile fits to the subdwarf O star BD+28$^\circ$4211. Two sets of
synthetic profiles are plotted, namely that from the pure H-He model structure
shown in Fig.\,\ref{tempfig} (dashed) and that from the model with CNO elements
and Stark broadened profiles. The Balmer line problem is occurring with the
former set and essentially disappears with the latter. Note in particular that
the H$\alpha$ emission core is perfectly matched.
}
\label{bd28fig}
\end{figure}

Fitting synthetic profiles to observed Balmer lines is the principal ingredient
of most spectroscopic analyses. The so-called Balmer line problem represents
the failure to achieve a consistent fit to the hydrogen Balmer line spectrum of
any hot sdO star whose effective temperature exceeds about 70\,000\,K. Results
of \Teff\ determinations drastically differ, up to a factor of two, depending
on which particular line is fitted. This problem was uncovered a few years ago 
during NLTE analyses of very hot subdwarfs and central stars of planetary
nebulae \cite{Napi94}. Since then, it cast severe doubt upon NLTE model
atmosphere analysis techniques as a whole. With new available models computed
with the ALI method we were able to demonstrate that the problem is due to the
neglect or improper inclusion of metal opacities \cite{W96}. We showed that the
Balmer line problem can be solved when surface cooling by photon escape from
the Stark wings of lines from the C, N, O elements is accounted for (see
Figs.\,\ref{tempfig} and \ref{bd28fig}).

\subsection{Heavy element abundances in sdO stars and white dwarfs}

\begin{figure}[bth]
\epsfxsize=9.8cm
\centerline{\epsffile{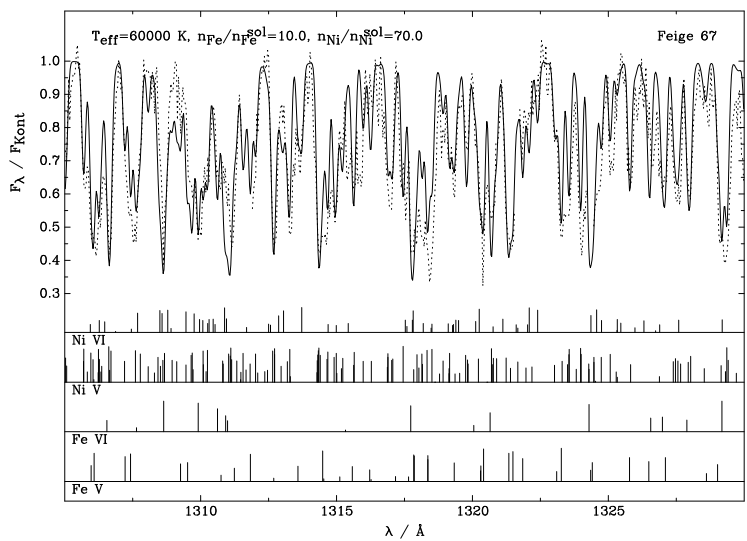}}
\caption{\small
Model fit (solid line) to the IUE spectrum of an sdO star. Iron and nickel
are overabundant by factors of 10 and 70, respectively, relative to solar
values. Vertical bars in the lower half of the panel indicate spectral line
positions of four different Fe and Ni ions and the bar heights correspond to
the relative $\log gf$ values (from \cite{Haas97}).
}
\label{feige67fig}
\end{figure}

The optical spectra of hot white dwarfs and sdO stars are dominated by helium
and/or hydrogen lines. Metals are highly ionized and their spectral lines are
almost exclusively located in the UV and extreme UV regions. High resolution
spectroscopy with the International Ultraviolet Explorer (IUE) has revealed a
wealth of spectral features from iron and nickel which, however, could not be
analyzed because of the lack of appropriate NLTE calculations. First attempts
for quantitative analyses were performed with line formation calculations on
pre-specified temperature and pressure model structures which in turn were
obtained from simplified LTE or NLTE models, i.e.\ disregarding metal line
blanketing effects \cite{Ven92,BB92}. Subsequently fully line blanketed LTE
models were employed, however, NLTE effects turned out to be non-negligible
\cite{DW92,DW93,WD94,LH95,Rau97}. Our latest models \cite{Haas96} include 1.5
million lines from the iron group elements, which are taken from Kurucz's
\cite{Kur91} line list. As an example for the quality of the fits we can
achieve, Fig.\,\ref{feige67fig} shows  a portion of the UV spectrum of the sdO
star Feige~67 and the best fitting model. The derived abundances suggest that
radiative levitation is responsible for the extraordinarily high heavy element
abundances in these stars.

\section{Conclusion}

We have described in detail the numerical solution of the classical model
atmosphere problem. The construction of metal line blanketed models in
hydrostatic and radiative equilibrium under NLTE conditions was the last and
long-standing problem of classical model atmosphere theory and it is finally
solved with a high degree of sophistication. Application of these models leads
to highly successful analyses of hot compact stars. Spectral properties from the
extreme UV through the optical region are for the first time  correctly
reproduced by these models. The essential milestones for this development,
starting from the pioneering work of Auer \& Mihalas \cite{AM69} are:

\begin{itemize}
\item Introduction of the Accelerated Lambda Iteration (ALI, or ``operator
splitting'' methods), based upon early work by Cannon \cite{Can73} and Scharmer
\cite{Schar81}. First ALI model atmospheres were constructed by Werner 
\cite{W86}.
\item Introduction of statistical approaches to treat the iron group elements
in NLTE by Anderson \cite{And89,And91}.
\item Computation of atomic data by Kurucz \cite{Kur91}, by the Opacity Project
\cite{Sea94} and subsequent improvements,
and by the Iron Project \cite{Hum93}.
\end{itemize}

\section{Acknowledgements}

We would like to thank Wolf-Rainer Hamann, Ulrich Heber, Ivan Hubeny and Thomas
Rauch for discussions, help, and contributions when we developed the {\tt PRO2}
code. We thank Ivan Hubeny for carefully reading the manuscript, which helped
to improve this paper. This work was funded during the recent years by the DFG
and DARA/DLR through several grants.

%
~
%

\end{document}